# High-Pressure Refractive Indices of NaCl, KCl, CaO, SrO, and MgO Reveal the Dependence of Anion Polarizability on Coordination Number and Bond Length

Konstantin Solovev[1,2], Xiangdong Li[1], Björn Winkler[3], Sergio Speziale[1], Sergey S. Lobanov[1,2]

[1]*GFZ Helmholtz Centre for Geosciences, Telegrafenberg, 14473 Potsdam, Germany*

[2]*Institute of Geosciences, University of Potsdam, Karl-Liebknecht-Str. 24-25, 14476 Potsdam Golm, Germany*

[3]*Institut für Geowissenschaften, Goethe-Universität Frankfurt, Altenhöferallee 1, 60438 Frankfurt am Main, Germany*

## ABSTRACT

Several alkali chlorides and alkaline-earth oxides adopt the B1 (rock-salt) structure at ambient conditions and transform to B2 (CsCl-type) under compression, allowing to change the coordination number and cation–anion distance independently across chemically distinct hosts. The change in anion polarizability under high pressure dominates the optical response of these prototypical ionic solids, yet its dependence on local structure and cation chemistry remains experimentally unconstrained at high pressure. Here we report refractive indices of NaCl and KCl to ~100 GPa, CaO to ~120 GPa, and SrO to ~60 GPa, each in both the B1 and B2 structures, complemented by published MgO-B1 indices to 140 GPa and first principles DFT calculations of the compounds in B1 and B2 structures. Lorentz–Lorenz analysis yields pressure-independent strain-polarizability parameters Λ ranging from 0.34±0.05 for KCl-B1 to 1.57±0.02 for MgO-B2. Our results establish coordination number and cation–anion distance as separable predictors of anion polarizability in both B1 and B2 structures. Increasing anion coordination from six to eight at fixed cation–anion distance lowers the polarizability of both $Cl^-$ and $O^{2-}$ by ~0.34–0.40 Å$^3$ and ~0.38–0.47 Å$^3$ respectively. Our results suggest that the polarizability of $O^{2-}$ is transferable to within ~0.06 Å$^3$ between MgO-B1 and CaO-B1 at fixed cation-anion distance, and that this transferability possibly extends to SrO-B1 and the B2 oxides. In contrast, the polarizability of $Cl^-$ is not transferable between NaCl-B2 and KCl-B2: the difference between them varies with cation-anion distance, from ~0.2 Å$^3$ at 2.65 Å to 0 at ~2.5 Å. The obtained anion polarizabilities show that the change in the strain polarizability parameter across the B1-B2 transition is governed by a decreasing sensitivity of the $O^{2-}$ polarizability to densification and, conversely, an increasing sensitivity of the $Cl^-$ polarizability to densification.

## I. INTRODUCTION

B1 (NaCl-type) and B2 (CsCl-type) crystals of alkali chlorides and alkaline-earth oxides are important model systems in condensed matter physics because they allow isolating fundamental structure-property relationships in simple ionic binary compounds. In particular, alkali halides and alkaline-earth oxides are regarded as prototypical ionic solids whose physical properties have long been of interest both for fundamental research and applications. At ambient conditions, sodium chloride (NaCl), potassium chloride (KCl), calcium oxide (CaO), strontium oxide (SrO), and magnesium oxide (MgO) are stable in the B1 crystal structure, with sixfold cation/anion coordination. All these compounds, however, undergo a pressure-induced phase transition to B2 crystal structures with eightfold cation/anion coordination at room temperature [1-7]: ~30 GPa (NaCl), ~2.0–2.6 GPa (KCl), ~60–66 GPa (CaO), and ~32–40 GPa (SrO) [1-7]. The transition of MgO-B1 to the B2 structure at room temperature has been observed at ~430-560 GPa in double-stage diamond anvil cell experiments [8], with *ab initio* calculations predicting the transition at 470(20) GPa [9]. First-order B1 to B2 phase transitions are among the most extensively studied transformations in high-pressure science, for their mechanism [10-18], mechanical properties [9,17,19-21], equations of state [3-5,14,22-26], and as benchmarks for computational methods, including machine learning, against high-quality experimental data [27,28]. The B1-B2 phase transition in CaO and MgO is also relevant to planetary sciences, because these oxides are major components of planetary mantles [14,29-31], where their physical and chemical properties may be sensitive to cation/anion coordination [14,32,33]. Because the B1-B2 phase transition can be induced by pressure alone, high-pressure studies on NaCl, KCl, CaO, SrO, and MgO offer an opportunity to isolate structural effects — via the coordination-number change at fixed composition — from chemical effects, accessible by comparing across these four compounds.

Refractive index gives access, through the Lorentz–Lorenz (LL) relation [34], to electronic polarizabilities thus providing insights into chemical bonding, and can be used to evaluate models of interionic interactions. For cubic crystals, the LL relation connects the macroscopic dielectric response to the electronic polarizability of each ion via a local-field correction. Because the electronic polarizability is particularly sensitive to the bonding environment, interatomic distance, and coordination, high-pressure refractive-index measurements over B1-B2 phase transitions offer a direct, quantitative route to probing how these variables govern the polarizability. In addition, the LL model yields a linear relationship between logarithms of polarizability and density over a wide pressure range (up to ~900 GPa and fourfold densification [35]), making it possible to relate a measured high-pressure refractive index directly to density and extrapolate refractive index to higher pressures. This relation is especially valuable in studies of non-crystalline materials, for which direct volumetric measurements at high pressures are challenging.

Early refractive-index measurements, conducted mostly at ambient pressure and extending to ~10 GPa for MgO, were used to test models of interionic interactions in alkali halides and alkaline-earth oxides [36-48]. These models initially treated the polarizability of crystals as a sum of independent free-ion polarizabilities. Subsequent studies, however, demonstrated this approximation to be insufficient, because the local ionic environment modifies each ion's electronic response. Such modified polarizabilities of ions in a crystal are often referred to as in-crystal polarizabilities. In particular, as was shown for cubic systems, in-crystal polarizabilities of closed-shell $s^2$ and $p^6$ cations are nearly equal to their free-ion values, whereas anion polarizabilities are significantly lower than their corresponding free-ion values [38,41,47-49]. It has also been shown, both experimentally and theoretically, that for highly ionic compounds such as NaCl, KCl, CaO, and MgO, polarizability can be separated into cation and anion contributions [47,50]. One unresolved problem is the pressure dependence of these individual contributions. The pressure dependence of the electronic polarizability of alkali and alkali earth cations in ionic compounds is expected to be weak [41,49,51,52]. Accordingly, the reduction of the anion polarizability under compression is the dominant change in the total electronic polarizability of highly ionic crystals, and hence in their refractive index. Therefore, establishing how the anion polarizability depends on the cation chemistry, cation-anion distance, and coordination potentially provides a basis for global models of refractive indices at high pressures [53]. Beyond being essential for such optical models, the anion polarizabilities are closely related to the dipole–dipole dispersion coefficients, which govern the leading term of the van der Waals attraction between two atoms [51,54,55]. Moreover, because anion polarizabilities are expected to be sensitive to the chemical environment and local structure [49,52,56], one can establish reference behaviors to use in studies at high pressures of solids lacking long range structural order (e.g., in compressed glasses). Last but not least, the refractive indices of NaCl, KCl, CaO, and MgO at high pressure are also of practical importance in static diamond anvil cell (DAC) and shockwave experiments, where they are often employed in the sample assembly. For instance, the refractive indices allow quantifying the sample thickness in DACs by white light interferometry, which is essential for *in situ* measurements of thermal and electrical conductivity [57,58].

Currently, except for MgO-B1 for which refractive-index data are available up to 140 GPa [59], the high-pressure refractive indices of NaCl, KCl, CaO, and SrO are mainly based on atomistic model calculations (e.g., [49,52,60]), with only limited experimental validation. Available experimental reports on the high-pressure refractive index are restricted to ~30 GPa for NaCl-B1 (shockwave experiments; [61,62]), to ~ 10 GPa for KCl-B1 and KCl-B2 (DAC; [63]). To the best of our knowledge, there are no experimental reports on the high-pressure refractive index of CaO-B1, CaO-B2, SrO-B1, and SrO-B2. These extant experimental data are too limited in pressure range and/or phase coverage to isolate the structural and chemical effects on anion polarizability as well as to establish a robust density-dependent equation for the polarizability, which requires index measurements over a broad compression range and across both B1 and B2 phases.

Here we report on the optical refractive indices of B1 and B2 phases of NaCl, KCl, CaO, and SrO up to 120 GPa, and combine our results with the previously reported indices of MgO-B1 up to 140 GPa. We supplemented the experimental data with our own density functional theory (DFT) computations of the indices of these phases as well as that of MgO-B2. Using the LL approach, we established a linear correlation between the logarithms of the polarizability and density of these materials. The slope of this correlation defines the strain-polarizability parameter ($\Lambda$), which we used to derive a density-dependent model of the refractive indices. The LL polarizabilities allowed us to establish the polarizabilities of $Cl^{-}$ and $O^{2-}$ at high pressure, and to isolate the effects of coordination number, bond length, and chemical environment on anion polarizability.

## II. THEORETICAL BACKGROUND AND METHODS

### A. Diamond anvil cell experiments

High-pressure experiments were performed using symmetric DACs equipped with matching type Ia diamonds with either flat (200 or 400 μm in diameter) or beveled (100 or 150 μm) culets. Rhenium gaskets with the initial thickness of approximately 250 μm were doubly indented between the anvils to a thickness of 10–15 μm and laser-drilled to form cylindrical sample chambers with diameters ~40% of the corresponding culet diameter. The refractive-index measurements described below require that the sample is optically homogeneous and fills the sample chamber entirely. To this end, we placed single crystals of NaCl and KCl of suitable dimensions (dried at 220 °C for 2 h) into the sample chamber, and sealed it immediately to minimize uptake of atmospheric moisture. Because CaO and SrO are very hygroscopic, we fired their powders in platinum crucibles at 1000-1200 °C and transferred hot into a dry nitrogen-purged glovebox for the DAC loading. Increasing the pressure to ~0.2 GPa for NaCl/KCl and to ~15 GPa for CaO and SrO was sufficient to make the samples optically homogeneous. In experiments with 200 and 400 μm culets, a ruby sphere was placed in the sample chamber to serve as a pressure gauge, with an assumed relative uncertainty of ±3% [64]. In experiments with 100 and 150 μm anvils, we determined the pressure from the shift of the high-frequency edge of the first-order diamond Raman band, with a relative uncertainty of ±5% [65].

**B. Refractive index measurements**

We measured the high-pressure refractive index using the optical reflectivity method [66,67], which we improved upon by using a supercontinuum laser as a probe [59,68-71]. The Fresnel equations relate the reflectance of the diamond–sample interface ($R_{dia-sam}$) to the real parts of the refractive indices of sample ($n_{sam}$) and diamond ($n_{dia}$). For normal incidence of light:

$$R_{dia-sam} = \left(\frac{n_{sam} - n_{dia}}{n_{sam} + n_{dia}}\right)^2 \quad (1)$$

The probe is inserted into the optical path of our DAC microscope by a non-polarizing beamsplitter cube after passing through a shortpass (950 nm cutoff) and a variable neutral density filter [70]. The use of a narrow, collimated laser beam (diameter of 1.2 mm at $\lambda = 440$ nm and 2.2 mm at $\lambda = 800$ nm) and a 20× Mitutoyo NIR objective (numerical aperture of 0.4) allows for a near-normal incidence of the probe and its small diameter at the focal plane (~5 µm). The reflected beam passes through a spatial filter (2 × 50 mm achromatic doublets and a 75 µm confocal pinhole), and is recorded on a Princeton Instruments PIXIS-100 CCD, installed on an IsoPlane 320 spectrometer (grating 600 g/mm, blaze 500 nm, wavelength calibration accuracy <0.2 nm).

For a thin transparent sample in the DAC, the measured reflected signal is a sum of partial reflections from the upstream and downstream diamond–sample interfaces. Accordingly, we analyzed the reflectance data using a two-beam interference model (Fig. 1), which allows accounting for ~ 99.9 % of the total reflected light reaching the detector [68]. To quantify $R_{dia-sam}$ (Eq. (1)), we collected a series of reflectance spectra from a reference mirror ($I_{mirror}$), the diamond–air interface ($I_{dia-air}$), and the diamond–sample interface ($I_{dia-sam}$) in the wavelength range of 550–650 nm at each pressure after a typical dwell time of ~24 h [68]. The intensity of the laser ($I_{laser}$) was determined from the measured reference mirror spectrum ($I_{mirror}$) and the known mirror reflectivity ($R_{mirror}$):

$$I_{laser} = \frac{I_{mirror}}{R_{mirror}} \quad (2)$$

The intensity of light impinging on the diamond–sample interface ($I_0$) was then calculated using $I_{laser}$ after accounting for reflection losses at the diamond-air interface ($I_{dia-sam}$):

$$I_0 = I_{laser}\left(1 - \frac{I_{dia-air}}{I_{laser}}\right) \quad (3)$$

Similarly, we obtained the intensity of light reflected from the diamond–sample interfaces ($I_1 + I_2$) from the spectrum measured off the sample ($I_{dia-sam}$) after accounting for the diamond-air reflection:

$$I_1 + I_2 = I_{dia-sam}\left(1 - \frac{I_{dia-air}}{I_{laser}}\right) \quad (4)$$

We then averaged the normalized spectrum ($\frac{I_1+I_2}{I_0}$) in the 550–650 nm range ($I_{avg}$), which is an excellent approximation of the phase average because of the weak index dispersion in this spectral range (< 1%), sufficient oscillation periods (> 10), and high data-point density (~1050 points). The two-beam interference model then relates $I_{avg}$ to $R_{dia-sam}$ [68]:

$$I_{avg} = R_{dia-sam}^3 - 2R_{dia-sam}^2 + 2R_{dia-sam} \quad (5)$$

Using the calculated $R_{dia-sam}$ obtained from the physically relevant roots of Eq. (5), we solved Eq. (1) for $n_{sam}$ under the assumption that $n_{dia}$=2.418 [72]. The validity of this assumption has been demonstrated at the level of ~1 % in the pressure range up to 140 GPa [59,67]. The obtained $n_{sam}$ is the effective refractive index in the range of 550-650 nm. Hereafter, we refer to it as the refractive index at 600 nm. We empirically estimated the reproducibility of the method as ±1% [59,68].

**C. Lorentz-Lorenz model and the density equation of refractive index**

Electromagnetic theory gives the refractive index of an insulator from the dielectric constant, which arises from induced dipoles in the material [73]. The Lorentz–Lorenz model accounts for local-field corrections, that is, the difference between the macroscopic applied electric field and the microscopic field acting on an individual polarizable unit [34,73,74]. This difference in the electric field arises from the contribution of surrounding induced dipoles to the local field. For isotropic materials, the refractive index is described by the following equation (CGS system):

$$\frac{n(\omega,\rho)^2 - 1}{n(\omega,\rho)^2 + 2} = \frac{4\pi}{3}\frac{\rho}{M} N_A \alpha(\omega,\rho) \quad (6)$$

where n and α denote refractive index and polarizability, respectively, both of which depend on frequency (ω) and density (ρ), M is the molar weight of the material, and $N_A$ is Avogadro's constant. For highly ionic compounds α can be described as a sum of polarizabilities of cations ($\alpha_{cat}$) and anions ($\alpha_{an}$) [47-49,52]:

$$\alpha = \alpha_{cat} + \alpha_{an} \quad (7)$$

The cation and anion polarizabilities in Eq. (7) are often referred to as in-crystal polarizabilities. The density equation of the polarizability can be obtained based on Eq. (6) by expressing α as a function of density at some frequency or wavelength (600 nm in this study) [75]:

$$\frac{d \ln \alpha(\rho)}{d \ln \rho} = -\Lambda(\rho) \quad (8)$$

where Λ, is a strain polarizability parameter. Generally, Λ is a positive number (because polarizability should decrease upon compression) that itself depends on density. Yet, experimental data show that for many materials over a wide range of achievable densifications (e.g., up to fourfold for LiF [35]), Λ is largely pressure independent. In that case, integrating Eq. (8) we obtain polarizability as a function of density:

$$\alpha(\rho) = \alpha_0 \left(\frac{\rho}{\rho_0}\right)^{-\Lambda} \quad (9)$$

where $\alpha_0$ and $\rho_0$ are the polarizability and density at a reference pressure (e.g., at 1 atm).

Using Eq. (6) and Eq. (9), the Lorentz–Lorenz equation can be written for arbitrary density (Eq. (10)):

$$L(\rho) = \frac{n(\rho)^2 - 1}{n(\rho)^2 + 2} = \frac{4\pi}{3}\frac{\rho}{M} N_A \alpha_0 \left(\frac{\rho}{\rho_0}\right)^{-\Lambda} \quad (10)$$

where L denotes the left-hand side of Eq. (6). We then normalize Eq. (10) by the value for a reference state $L(\rho_0)$ to eliminate the density and polarizability at the reference density, yielding Eq. (11), in which the refractive index depends only on the strain polarizability parameter Λ and densification ratio:

$$\frac{L(\rho)}{L(\rho_0)} = \frac{\rho}{\rho_0}\left(\frac{\rho}{\rho_0}\right)^{-\Lambda} = \left(\frac{\rho}{\rho_0}\right)^{1-\Lambda} \quad (11)$$

Eq. (11) represents the density equation for refractive index at high pressures. Refractive index at high pressures can then be directly calculated from L(ρ) using refractive index at reference density (in our case, the value at 1 atm, $L(\rho_0)$) and densification ratio (e.g., obtained from appropriate EOS).

Assuming that cations polarizability is density independent, in agreement with current theories [38,47,49,52,56], we substitute Eq. (7) into Eq. (8), yielding Eq. (12):

$$\frac{d\ln(\alpha_{cat} + \alpha_{an}(\rho))}{d\ln\rho} = \frac{\alpha_{an}(\rho)}{\alpha_{cat} + \alpha_{an}(\rho)} \frac{d\ln\alpha_{an}(\rho)}{d\ln\rho} = -\Lambda(\rho) \quad (12)$$

Equation (12) shows that the total logarithmic derivative is the anion derivative weighted by the fractional anion contribution. Consequently, a density-independent cation contribution reduces the magnitude of the total derivative, whereas the magnitude and density dependence of the anion polarizability determines the high-pressure evolution of the refractive index.

**C. Density functional theory (DFT) computations**

First principles calculations were carried out within the framework of density functional theory (DFT) using commercial and academic versions of the CASTEP program [76,77]. The calculations were performed employing the Perdew-Burke-Ernzerhof generalized gradient approximation GGA-PBE [78] and a plane wave basis set in conjunction with pseudopotentials from the CASTEP database. "On the fly" ultrasoft pseudopotentials generated using the descriptors in the CASTEP data base were employed in conjunction with plane waves up to a maximum cutoff energy of 570 eV. The accuracy of the pseudopotentials is well established [79]. Monkhorst-Pack grids [80] for sampling of the reciprocal space were chose so that k-point separations of less than 0.025Å-1 resulted (e.g., 18 x 18 x 18 for MgO-B2 at 20 GPa). The calculations were considered to be converged once the maximal residual force acting on an atom was < 0.01 eV/Å, the residual stress was < 0.02 GPa, and the maximal energy change was < 5 x 10-6 eV/atom.

It is well established that DFT-GGA-PBE calculations underestimate the band gap substantially, and that this will result in an overestimation of refractive indices. This systematic off-set was generally found to be small. However, for CaO the discrepancies between experiment and DFT were larger and hence we carried out additional calculations using the HSE06-functional for the calculation of optical properties. While typically HSE06-calculations result in band gaps which are in much better agreement with experiment [81-83], they are computationally significantly more expensive than DFT-GGA-PBE calculations, so a coarser Monkhorst-Pack grid had to be chosen (e.g. 8 x 8 x 8 for CaO-B2 at 20 GPa). Also, norm-conserving pseudopotentials with significantly higher cut-offs (e.g., 1020 eV for CaO) are required. These calculations gave a refractive index of 1.757 at ambient conditions, much lower than the 2.2262 for a GGA-PBE calculation, but the pressure dependence of the refractive index is essentially the same for both approaches. Hence, in the analysis we only used the DFT-GGA-PBE results.

## III. RESULTS AND DISCUSSION

### A. High-pressure refractive indices of NaCl, KCl, and CaO

Upon compression, the refractive index of NaCl-B1 increased by about 10%, from 1.5434(1) at 1 atm at 600 nm [84] to 1.699(17) at 28.8(9) GPa (Fig. 2a). From 1 atm to about 10 GPa, the pressure derivative decreased, and above 10 GPa it remained nearly constant up to 28 GPa. At ~28–35 GPa, which corresponds to the pressure range of the B1-B2 phase transition [22], the optical quality of the NaCl sample deteriorated, presumably due light scattering at grain boundaries of the coexisting B1 and B2 NaCl phases. For this reason, on compression we measured the index of NaCl-B2 only at $P > 35(1)$ GPa (after the completion of the B1-> B2 transition) and up to 101(5) GPa. In this pressure range, the index increased by ~7.5%, from 1.723(17) to 1.852(18), with the slope gradually decreasing with pressure (Fig. 2a). Upon decompression, the refractive index of NaCl was indistinguishable from that obtained on compression. Interestingly, NaCl-B2 could be preserved upon decompression down to 24.8(7) GPa, at which pressure we observed both B1 and B2 phases coexisting in separate regions of the sample chamber (Fig. 2a inset) and could thus measure their indices at the same DAC load (1.662(17) for NaCl-B1 and 1.715(17) for NaCl-B2 at 24.8(7) GPa). The refractive index of KCl-B1 increased approximately linearly by about 2.4%, from 1.4896(1) at 1 atm at 600 nm [84] to 1.525(15) at 2.33(7) GPa (Fig. 2a and Fig. 2b).The phase transition to KCl-B2 was observed visually at 2.33(7) – 2.58(8) GPa, consistent with the literature [3]. The index of KCl-B2 at 3.4(1) GPa is 1.620(16) and increases further with pressure at a gradually decreasing rate up to ~40 GPa. Above this pressure, the rate of increase in index is nearly pressure independent. At the highest experimental pressure of 97(5) GPa the refractive index of KCl-B2 was 2.00(2), corresponding to a total increase of about 23.5% relative to the first measured pressure point at 3.4(1) GPa. Similarly, to NaCl, the refractive index of KCl measured on decompression is fully consistent with that measured on compression. The refractive indices of KCl reported previously up to ~10 GPa [63] agree broadly with our results but are systematically ~0.5-1% lower at corresponding pressures than that measured here (Fig. 2b). Our experimental data for NaCl-B1, NaCl-B2, KCl-B1, and KCl-B2 are also in good agreement with the results of our DFT-PBE calculations (Fig. 2a and Fig. 2b). Specifically, the overall increase in the refractive index obtained by DFT-PBE calculations over corresponding densities is close to that obtained from experimental data (e.g., 6.7% for NaCl-B2, and 25.8% for KCl-B2). In addition to reproducing the relative change in the refractive index, the calculations capture the changes in slope over specific pressure ranges. We note, however, that the calculated values are systematically higher than the experimental ones by ~1–5 %, likely because DFT-PBE tends to underestimate the band gap.

We measured the refractive index of CaO-B1 up to 62(3) GPa, corresponding to the onset of the B1-> B2 phase transition [4], and then the index of CaO-B2 up to 120(6) GPa (Fig. 2c). The refractive index of CaO-B1 increases modestly upon compression from 1.837(1) at 1 atm at 600 nm [85] to 1.893(19) at 62(3) GPa (Fig. 2c). The refractive index of CaO-B2 also increases with pressure from 2.00(2) at 72(4) GPa to 2.06(2) at 120(6) GPa (Fig. 2c). The *ab initio* computed indices of CaO-B1 and CaO-B2 are ~10% (at 1 atm) and ~16 % (at 72 GPa) higher than that measured experimentally. The slope of CaO-B1 obtained by DFT-PBE is similar to that obtained in experiments. Yet, the calculated refractive index of CaO-B2 increases noticeably faster with pressure than observed experimentally. Specifically, the refractive index computed at the densities corresponding to the first and last experimental data points yields an increase of 6.9%, compared to 3.3% from the experimental results. Although the overall increase of the refractive index of CaO is smaller than that of NaCl and KCl, it remains positive in the studied pressure range. The refractive index of SrO-B1 increases slightly, from 1.8695(1) at 1 atm at 600 nm [86] to 1.89(2) at 32(2) GPa. We observed the B1-to-B2 phase transition in SrO at 32(2) GPa, consistent with previously reported experimental transition pressures of approximately 32–40 GPa [10]. After the transition, the refractive index of SrO-B2 was 1.98(2) at 43(2) GPa and increased by ~2% to 2.02(2) at 60(3) GPa. During decompression from 38(2) to 25(1) GPa, the B1 and B2 phases coexisted, allowing their refractive indices to be measured at the same DAC loads. In contrast to B1 and B2 phases of CaO and SrO, the refractive index of MgO-B1 reported by Schifferle, et al. [59] decreases upon compression (Fig. 2c).

All the refractive indices reported here, as well as that of MgO-B1 reported previously [59], can be accessed interactively via https://glass2melt.info/. The website allows visualization of the experimentally obtained refractive index as a function of pressure. This can be useful for the design and interpretation of high-pressure optical experiments, as well as for educational and modelling purposes.

### B. Electronic polarizability of statically compressed NaCl, KCl, CaO, MgO, and SrO

Figure 3 shows the Lorentz-Lorenz (LL) polarizabilities of NaCl, KCl, CaO, SrO, and MgO obtained using Eq. 6 from the measured refractive indices and density data either from published EoS [3-6,14,22,25,87] or from our own DFT-PBE calculations. The parameters of the EOS of CaO-B1 ($K_{T0}$ = 122(2) GPa, $K'_{T0}$ = 3.6(1), $V_0$ = 111.39(6) Å$^3$) and CaO-B2 ($K_{T0}$ = 128(50) GPa, $K'_{T0}$ = 3.5(6), $V_0$ = 24.7(2) Å$^3$) were obtained by fitting the third-order Birch–Murnaghan equation to data from [4,5,14,25]. The uncertainties in the densities of the B1 and B2 phases of NaCl and KCl, and B1 phases of CaO and MgO obtained from the EoS are primarily determined by the 5% uncertainty in our experimental pressure, resulting in density uncertainties of 0.1–1.1% across these phases. In contrast, the EoS parameters for CaO-B2, SrO-B1, and SrO-B2 have larger uncertainties, resulting in total density uncertainties of ~2% for CaO-B2, 1.0–1.6% for SrO-B1, and 1.5–2.5% for SrO-B2. The propagation of uncertainties in the density and the refractive index (±1%) results in an overall relative uncertainty of approximately ±2% in the LL polarizability for all compounds except CaO-B2 and SrO, for which the larger uncertainty in the density results in an uncertainty of approximately ±2–4% (Fig. 3). The LL polarizability decreases monotonically with increasing pressure for all the investigated compounds, regardless of crystal structure. The decrease is strongest at low pressure and gradually becomes weaker at higher pressure, typical of many other materials investigated at high pressure [67,88-90]. We will use the LL polarizabilities in Section C to construct the polarizability-based density model of the refractive index.

### C. Density dependence of the electronic polarizability

We can now fit equation (9) to the LL polarizabilities for B1 and B2 phases of NaCl, KCl, CaO, SrO and MgO to obtain $\Lambda$, the strain polarizability parameter, assuming it is density independent, which we discuss below. In the absence of experimental data on MgO-B2, we fitted equation (9) to LL polarizability inferred from our own DFT-PBE refractive index computations. Figure 4 shows the LL polarizabilities normalized either to the value at 1 atm (B1 phases) or to the lowest-pressure value available to us (B2 phases). For phases stable at 1 atm, the linear fit was constrained to have zero intercept, that is the starting density was fixed at the experimental value. The obtained values are listed in Table I. Please note that values of $\Lambda$ greater than 1 imply a negative density dependence of the refractive index (Eq. 12), as observed in the case of MgO-B1 [45,91], but also for $Al_2O_3$ [92] and stishovite [68]. The values of $\Lambda = 0.34(5)$ for KCl-B1 and of $\Lambda = 0.42(2)$ for KCl-B2 are different from those we obtained from the LL polarizabilities based on the previously reported refractive index data for statically compressed KCl-B1 ($\Lambda = 0.46(3)$; [63]) and KCl-B2 ($\Lambda = 0.36(4)$, [63]), and the same EoS [3,22] we used with our own index data. We assign the discrepancy in the B1 phase of KCl to the ~1% relative error in index associated with our own measurements that make it difficult to reliably resolve the density slope of the refractive index in the narrow pressure range of KCl-B1 stability (< ~3 GPa [3]). For the B2 phase of KCl, the discrepancy likely arises from the very limited density range covered by the literature refractive index data (≲9 GPa), which hinders a reliable determination of the density dependence of the refractive index. In addition, the strain polarizability parameter for MgO-B1 obtained here $\Lambda = 1.14(2)$ is lower than the values we obtained using previously reported indices: $\Lambda = 1.28$ [45] and $\Lambda = 1.23$–$1.27$ [91], from studies restricted to 10 and 0.7 GPa, respectively. In the case of MgO, we suppose that it is the previous studies that did not reliably capture the density dependence of the refractive index of stiff MgO at small densifications (up to ~6 %), as opposed to the work of Schifferle, et al. [59], whose index data allowed us to fit $\Lambda$ over the ~45% densification (Fig. 4i).

Using the strain polarizability parameters in Eq. 12, we modeled the refractive index of all the studied materials (Fig. 5). For materials stable at 1 atm (i.e., all B1 phases of NaCl, KCl, CaO, and MgO), we used their known densities and indices at 1 atm, and their polarizabilities from Eq. (6), to calculate $L(\rho_0)$. For materials unstable at 1 atm (i.e., all B2 phases), we use the density at 1 atm obtained from their reported equations of state (EOS), the extrapolated polarizability from Eq. (10), and the refractive index calculated from these two values using Eq. (6), to determine $L(\rho_0)$. Figure 5 shows that the Lorentz–Lorenz model provides an excellent description of the static-compression data over the full pressure range examined in the experiments.

The predictive power of Eq. (11), however, may be limited by the assumption that the strain polarizability parameter remains constant over a wide pressure range. To test this assumption, we used exclusively our DFT-PBE results for NaCl-B2, KCl-B2, and MgO-B1. Specifically, we tested whether a $\Lambda$ obtained from DFT over a restricted range of densification, chosen to match that covered by our experimental measurements, reproduces the DFT values of $\ln(\alpha/\alpha_0)$ at higher densities. Figure 6 shows that for DFT-PBE a constant strain polarizability parameter underestimates $\ln(\alpha/\alpha_0)$ beyond the fitting range, with this error increasing progressively toward higher degrees of densification. For NaCl-B2, this error corresponds to an underestimation of its refractive index by ~0.3–4.4% at 110–300 GPa ($\rho/\rho_0 \sim 2.1$–$2.9$). For KCl-B2, the modelled refractive index would be underestimated by ~1–17% at 110–300 GPa ($\rho/\rho_0 \sim 2.3$–$2.5$). For MgO-B1, the modelled refractive index would be underestimated by about 1–3% at 200–300 GPa ($\rho/\rho_0 \sim 1.6$–$1.8$). This confirms the inference that the extrapolation error associated with the pressure-independent strain polarizability approximation is small for less compressible materials in the same family of compounds (i.e., sharing the same anion). At the same time, the assumption of a pressure-independent $\Lambda$ appears to remain valid for the B1 phase of LiF (bulk modulus of ~66 GPa at 1 atm [23]), yielding good agreement with shock-wave refractive index data up to 900 GPa, corresponding to more than fourfold densification [35].

**D. Structural Effects on Anion Polarizability**

Despite the importance of ionic polarizabilities, only limited experimental data are available [53], complemented mainly by theoretical calculations [38,41,47-49,52,56]. We address this gap by extracting the anion polarizabilities from the LL polarizabilities of the studied phases assuming additive behavior of cation and anion polarizabilities (in-crystal polarizabilities; Eq. (7)). The validity of this assumption for ionic crystals is supported by theoretical arguments as well as *ab initio* calculations [38,47,49,52,56]. These studies have shown that for closed-shell s/p cations, such as $Na^+$, $K^+$, $Mg^{2+}$, $Ca^{2+}$, $Sr^{2+}$ (and $Ba^{2+}$), the cation contribution is nearly equal to the free-ion value, depends only weakly on the surrounding environment, and is largely density independent [48,49]. In contrast, anion polarizabilities are sensitive to the interatomic distance and crystal structure [47-49,51,52]. Accordingly, we calculated the polarizabilities of $Cl^-$ and $O^{2-}$ by subtracting the corresponding free-ion cation polarizabilities (Table II) from the total LL polarizability, assuming the cation polarizabilities are pressure-independent. For comparison, we added also the LL polarizability of BaO-B1 at 1 atm, which we obtained using Eq. (6) from the refractive index of BaO [93] and its density [94]. Figure 7 shows the resulting anion polarizabilities as a function of the cation–anion separation. Here we chose this variable because the anion polarizability is mainly affected by electrostatic interactions and nearest-neighbor overlap [48,49,51,52].

First, we isolate the effect of cation coordination on the anion polarizability. At a fixed $Cation^+$–$Cl^-$ distance, the electronic polarizability of $Cl^-$ in the B2 phase is lower than that in the B1 phase. That is, increasing the coordination number from six to eight while keeping the $Cation^+$–$Cl^-$ distance constant reduces the polarizability of $Cl^-$ by ~0.34–0.40 Å$^3$ (Fig 7a). Similarly, at a fixed $Cation^+$–$O^{2-}$ separation, the polarizability of $O^{2-}$ in CaO-B2 and SrO-B2 is ~0.38-0.47 Å$^3$ lower than in the corresponding B1 phases (Fig 7b). Thus, in both the chlorides and the oxides, increasing the coordination number from six to eight across the B1–B2 transition lowers the anion polarizability at fixed cation-anion distance. Nearest-neighbor interactions therefore dominate the evolution of anion polarizabilities in these systems, as also discussed in previous works based on *ab initio* calculations (e.g., [52]).

To explore the effect of different chemical environments, we compared the polarizabilities of $Cl^-$ in NaCl-B2 and KCl-B2 as well as $O^{2-}$ in CaO-B1 and SrO-B1 at the same cation-anion distance. The polarizability of $Cl^-$ in KCl-B2 is lower than that in NaCl-B2 over a broad range of $Cation^+$–$Cl^-$ separations and decreases from ~0.20 Å$^3$ at ~2.65 Å to less than ~0.10 Å$^3$ at a distance of 2.55 Å. $O^{2-}$ polarizability is ~0.10 Å$^3$ lower in SrO-B1 than in CaO-B1 at ~2.4 Å, where data comparison for the same distances is possible. Thus, in both the chlorides and the oxides, the anion is less polarizable in the presence of larger cations at a fixed cation-anion distance. This behavior would be expected because the greater spatial extent of the electron density of the larger cations will, through the operation of the Pauli principle, exert a greater compressive influence on the neighboring anion electron density, reducing its polarizability. Although our experimental data offer such comparisons only in a relatively narrow range of metal–anion separations and not across all studied compounds, our own DFT-PBE predictions as well as previous *ab initio* computations [49,52] support the available observations, and suggest that the identified behavior is systematic, extending reliably to much broader ranges of cation-anion distances. The same type of dependence on coordination, cation-anion separation, and chemical environment (type of cation), were obtained by *ab initio* calculations for the $I^-$ ion in NaI and KI [56].

**E. Are Anion Polarizabilities Transferable?**

Previous computational studies found the polarizabilities of $F^-$ [48,52], $Cl^-$ and $Br^-$ [52] to be partially transferable in compounds with different cations. Specifically, the shape of the anion polarizability as a function of cation–anion distance is similar for a given anion, but displaced by a cation-dependent offset (for $Cl^-$, the offset is ~0.30 $Å^3$ at both 2.55 and 2.65 Å according to the NaCl-B1 and KCl-B1 curves by Wilson, et al. [49]; values from Jemmer, et al. [52] are ~0.36 $Å^3$ at 2.55 Å and ~0.43 $Å^3$ at 2.65 Å). Our measurements test these predictions under high pressure over cation–$Cl^-$ distances of ~2.4-3.1 Å (Fig. 7a). In the B2 structure, where NaCl and KCl overlap in cation–$Cl^-$ distance, the polarizabilities of $Cl^-$ are separated by ~0.20 $Å^3$ at ~2.65 Å. This is significant against a propagated experimental uncertainty in anion polarizability at high pressure of ~0.10 $Å^3$. The separation in $Cl^-$ polarizability decreases with decreasing cation-anion distance, with a crossing indicated at ~2.5 Å both in our experimental and DFT-PBE data. Previous theoretical studies did not observe such a crossing in $Cl^-$ polarizabilities [49,52]. Such a direct comparison for B1-structured NaCl and KCl is not possible because these do not overlap in cation–anion distance. However, extrapolating to a common distance along the MP2 curve computed for NaCl-B1, the polarizability of $Cl^-$ in KCl-B1 is ~0.2-0.3 $Å^3$ lower than that in NaCl-B1 (Fig. 7a). Our results therefore confirm the existence of a cation-dependent offset in $Cl^-$ polarizability predicted previously [49,52], but show that its magnitude and sign depend on the cation-$Cl^-$ distance rather than being constant. Accordingly, we find that the polarizability of $Cl^-$ is not transferable between NaCl and KCl.

The polarizabilities of $O^{2-}$ in MgO, CaO, and SrO at 1 atm, all derived from literature data, are within ~0.06 $Å^3$ of the CLUS (SCF) curve computed for MgO-B1 by Jemmer, et al. [52] (Fig. 7b). This is remarkable given that the cation polarizabilities we subtracted from the LL polarizabilities span ~1.4 $Å^3$ (0.0720, 0.4731, and 0.7706 $Å^3$ for $Mg^{2+}$, $Ca^{2+}$, and $Sr^{2+}$), ~13 times the remaining scatter of ~0.06 $Å^3$. Our high-pressure data for B1 phases of MgO and CaO also lie within ~0.06 $Å^3$ of the CLUS curve for MgO-B1, whereas all SrO-B1 data fall ~0.20 $Å^3$ below the CLUS curve. The in-crystal polarizabilities for $Sr^{2+}$ and $Ba^{2+}$ reported by Fowler and Pyper [48] were obtained by combining experimental refractive index data with accurate *ab initio* calculations for $Mg^{2+}$, $Ca^{2+}$, $Li^+$, $Na^+$, and $K^+$, and fitting the environmental dependence of multiple anions (oxides, sulphides, and fluorides) against interionic separation. We expect that this approach accumulates error in the extracted values of polarizability of $Sr^{2+}$ and $Ba^{2+}$, because these are obtained by difference, after subtracting anion polarizability derived from the empirical polynomials extrapolated to larger cation-anion distances in SrO, SrS, $SrF_2$, BaO, BaS, and $BaF_2$. The larger deviation for SrO is therefore consistent with the accumulated uncertainty in $Sr^{2+}$ polarizability rather than with a breakdown of transferability for larger cations. Our high-pressure data indicate that the polarizability of $O^{2-}$ is transferable across B1-oxides at the investigated cation-$O^{2-}$ distances. Our measurements on CaO-B2 and SrO-B2 suggest that $O^{2-}$ polarizability may also be transferable for B2-structured oxides.

**F. Structural Effects on the Strain Polarizability Parameter**

The calculated polarizabilities of $Cl^-$ and $O^{2-}$ enable us to directly investigate the structural dependence of the strain polarizability parameter $\Lambda$ (Eq. (8)) through the structural dependence of the anion polarizability, expressed by Eq. (12). For the studied materials, the $a_{an}/(a_{cat} + a_{an})$ ratios across the full pressure range are summarized in Table III. Densification causes this ratio to decrease as the $a_{cat}/a_{an}$ ratio increases. Combining these ratios with Eq. (12) yields the logarithmic density derivative of the anion polarizability, summarized in Table IV. Upon transitioning from the B1 to the B2 phase, the decrease in the logarithmic density derivative indicates that the $O^{2-}$ polarizability becomes moderately less sensitive to densification. Conversely, the B1-to-B2 transition in chlorides is accompanied by an increase in the sensitivity of the $Cl^-$ polarizability to densification. Within each chemical system, the $a_{an}/(a_{cat} + a_{an})$ ratio varies only marginally across the B1–B2 transition, with values immediately below and above the phase boundary differing by less than ~2%. In contrast, the density derivatives of the anion polarizability exhibit substantially larger discontinuities, changing by approximately ~8–10% for NaCl, ~27–37% for KCl, ~13–20% for CaO, and ~27–31% for SrO. This is why the magnitude and sign of the change in Λ across the B1–B2 transition are primarily governed by alterations in the response mechanism of the anion polarizability upon densification. Nevertheless, the cation contribution remains critical in determining Λ. For instance, although the absolute logarithmic density derivative of the $O^{2-}$ polarizability is higher in SrO-B1 than in MgO-B1 (Table IV), the negligible polarizability of $Mg^{2+}$ ultimately drives a decrease in the refractive index of MgO-B1 upon compression, as the overall polarizability is dominated by the anion contribution.

## IV. CONCLUSIONS

Here we combined experiments and first-principles computations to report on the refractive index of B1 and B2 phases of NaCl, KCl, CaO, SrO, and MgO at high pressure. Using these in the Lorentz–Lorenz equation, we developed the density models for the refractive index based on the polarizability, which was determined using the following pressure-independent strain polarizability parameters: Λ = 0.56±0.01 for NaCl-B1, Λ = 0.61±0.02 for NaCl-B2, Λ = 0.34±0.05 for KCl-B1, Λ = 0.42±0.02 for KCl-B2, Λ = 0.87±0.02 for CaO-B1, Λ = 0.69±0.03 for CaO-B2, Λ = 0.98±0.03 for SrO-B1, Λ = 0.6±0.1 for SrO-B2, Λ = 1.14±0.02 for MgO-B1, Λ = 1.565±0.023 for MgO-B2 (Table I). Polarizability-based refractive index models can be extended to other compounds and crystal structures. The Lorentz–Lorenz approach presented here will aid the development of high-pressure refractive index models for other transparent ionic solids. Another outcome of this study is the identification of systematic trends in $Cl^-$ and $O^{2-}$ polarizabilities across the B1 and B2 phases. Using the additivity of ionic polarizabilities, we isolated the anion contribution from the total Lorentz–Lorenz polarizability. At fixed catio–anion distance, increasing coordination from six to eight lowers the $Cl^-$ polarizability by ~0.34–0.40 Å³ in NaCl and KCl and the $O^{2-}$ polarizability by ~0.38–0.47 Å³ in CaO and SrO. Cation identity also matters: $Cl^-$ is ~0.10–0.20 Å³ less polarizable in KCl-B2 than in NaCl-B2, while $O^{2-}$ is ~0.10 Å³ less polarizable in SrO-B1 than in CaO-B1. These trends agree with our DFT-PBE results and previous *ab initio* calculations. The polarizability of $O^{2-}$ remains comparatively transferable within alkaline-earth oxides, whereas $Cl^-$ shows stronger cation dependence. Therefore, cation-anion distance, coordination, and cation chemistry are useful descriptors of anion polarizabilities and can be used to estimate refractive index of B1 and B2 compounds from structural information. From the obtained anion polarizabilities, we conclude that the change in the strain polarizability parameter Λ across the B1-B2 transition is governed by a decreasing sensitivity of the $O^{2-}$ polarizability to densification in oxides and, conversely, an increasing sensitivity of the $Cl^-$ to densification in chlorides.

## ACKNOWLEDGMENTS

This work was supported by the Helmholtz Young Investigators Group CLEAR (VH-NG-1325). K.S., X.L., S.S.L and S.S. were funded by the European Union (ERC, GLASS2MELT, 101126078). B.W. is grateful for support by the Dassault Systémes Science Ambassador program.

Table I. The strain polarizability parameters (Λ) of B1 and B2 phases of NaCl, KCl, CaO, SrO and MgO. The strain polarizability parameter of MgO-B2 was calculate based on DFT-PBE calculations (in the pressure range from 0 to 150 GPa). The uncertainties here represent 95% confidence intervals.

| | NaCl | KCl | CaO | SrO | MgO |
|---|---|---|---|---|---|
| B1 phase | 0.56(1) | 0.34(5) | 0.87(2) | 0.98(3) | 1.14(2) |
| B2 phase | 0.61(2) | 0.42(2) | 0.69(3) | 0.6(1) | 1.565(23) |

Table II. Free-ion cation polarizabilities for optical frequencies of $Na^+$, $K^+$, $Mg^{2+}$, $Ca^{2+}$, $Sr^{2+}$, and $Ba^{2+}$ from Fowler and Pyper [48].

| | $Na^+$ | $K^+$ | $Mg^{2+}$ | $Ca^{2+}$ | $Sr^{2+}$ | $Ba^{2+}$ |
|---|---|---|---|---|---|---|
| Cation polarizability | 0.1485 Å$^3$ | 0.7912 Å$^3$ | 0.0720 Å$^3$ | 0.4731 Å$^3$ | 0.7706 Å$^3$ | 1.4967 Å$^3$ |

Table III. The ratio $a_{an}/(a_{cat} + a_{an})$ of B1 and B2 phases of NaCl, KCl, CaO, SrO and MgO over the measured pressure range (Eq. (12)). The values are reported from the low- to high-pressure direction. The refractive index data of MgO-B1 is from Schifferle, et al. [59].

| | NaCl | KCl | CaO | SrO | MgO |
|---|---|---|---|---|---|
| B1 phase | 0.956(1)–<br>0.944(1) | 0.812(4)–<br>0.808(4) | 0.839(4)–<br>0.793(3) | 0.793(4)–<br>0.743(4) | 0.960(1)–<br>0.938(1) |
| B2 phase | 0.946(1)–<br>0.934(1) | 0.803(4)–<br>0.742(4) | 0.778(2)–<br>0.760(2) | 0.730(5)–<br>0.714(5) | |

Table IV. The logarithmic density derivative of the anion polarizability of NaCl, KCl, CaO, SrO and MgO over the measured pressure range (Eq. (12)). The results are reported as absolute values, although the corresponding derivatives are negative. The values are reported from the low- to high-pressure direction. The refractive index data of MgO-B1 is from Schifferle, et al. [59].

| | NaCl | KCl | CaO | SrO | MgO |
|---|---|---|---|---|---|
| B1 phase | 0.59(2) | 0.41(6) | 1.04(4)–<br>1.10(4) | 1.23(5)–<br>1.31(5) | 1.18(2)–<br>1.21(2) |
| B2 phase | 0.64(3)–<br>0.65(3) | 0.52(3)–<br>0.56(3) | 0.88(5)–<br>0.90(5) | 0.9(1) | |

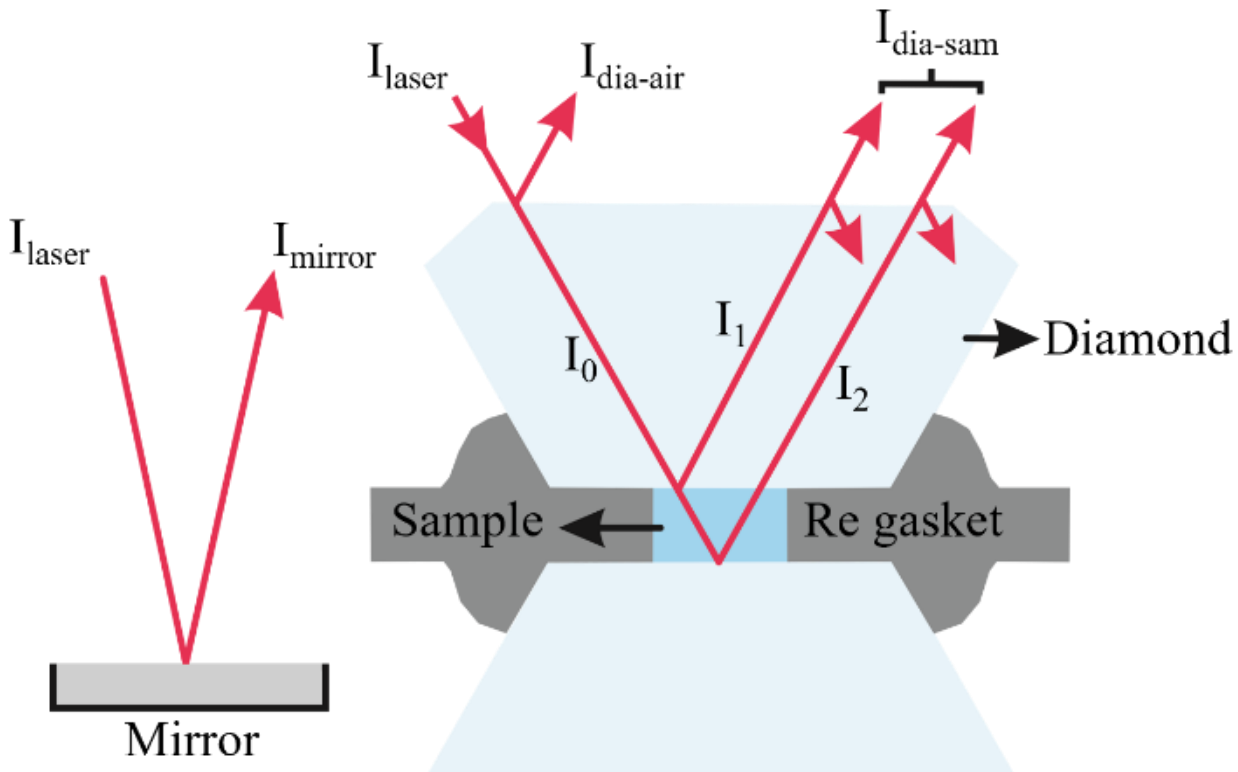

FIG. 1. Schematic diagram of the reflectivity measurements in the diamond anvil cell used to obtain the high-pressure refractive index. $I_{laser}$ is the laser intensity, determined from by measuring the intensity reflected off the reference mirror ($I_{mirror}$) with a known reflectance. $I_0$ is the light incident on the diamond–sample interface, obtained by correcting $I_{laser}$ for reflection at the upstream diamond–air interface. $I_1$ and $I_2$ are the individual reflections from the upstream and downstream diamond–sample interfaces. The total reflectance spectrum ($I_1$ + $I_2$) is obtained by correcting the measured $I_{dia–sam}$ spectrum for reflection losses at the upstream diamond–air interface using $I_{dia–air}$ and $I_{laser}$.

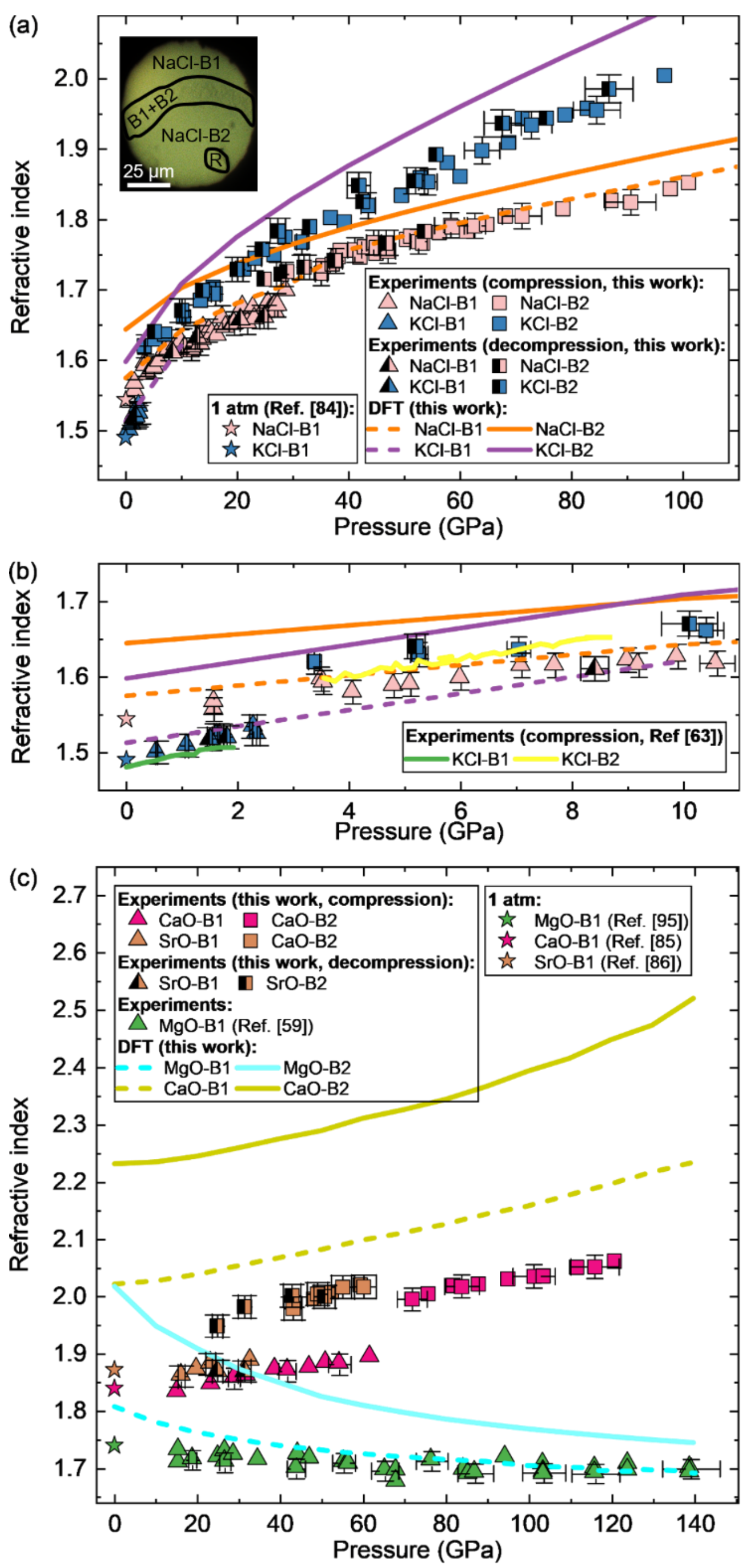

FIG. 2. Room-temperature high-pressure refractive indices of NaCl and KCl (a) and (b) as well as CaO, SrO, and MgO (from [59]) (c) measured in diamond anvil cell experiments at 600 nm in this work. Our DFT-PBE computations are also show by colored curves (see legend). The 1 atm values are from [84-86,95]. Subfigure (b) shows an enlarged region of subfigure (a) and therefore has the same legends, with the addition of KCl data from [63]. Inset in subfigure (a): Optical image of the NaCl sample at 24(1) GPa, where the B1 and B2 phases coexist (R denotes the ruby sphere). Only some error bars are shown because of the high data density. The reproducibility error for the refractive index is ±1%. The pressure reproducibility error is ±3% or ±5% depending on the pressure measurement method.

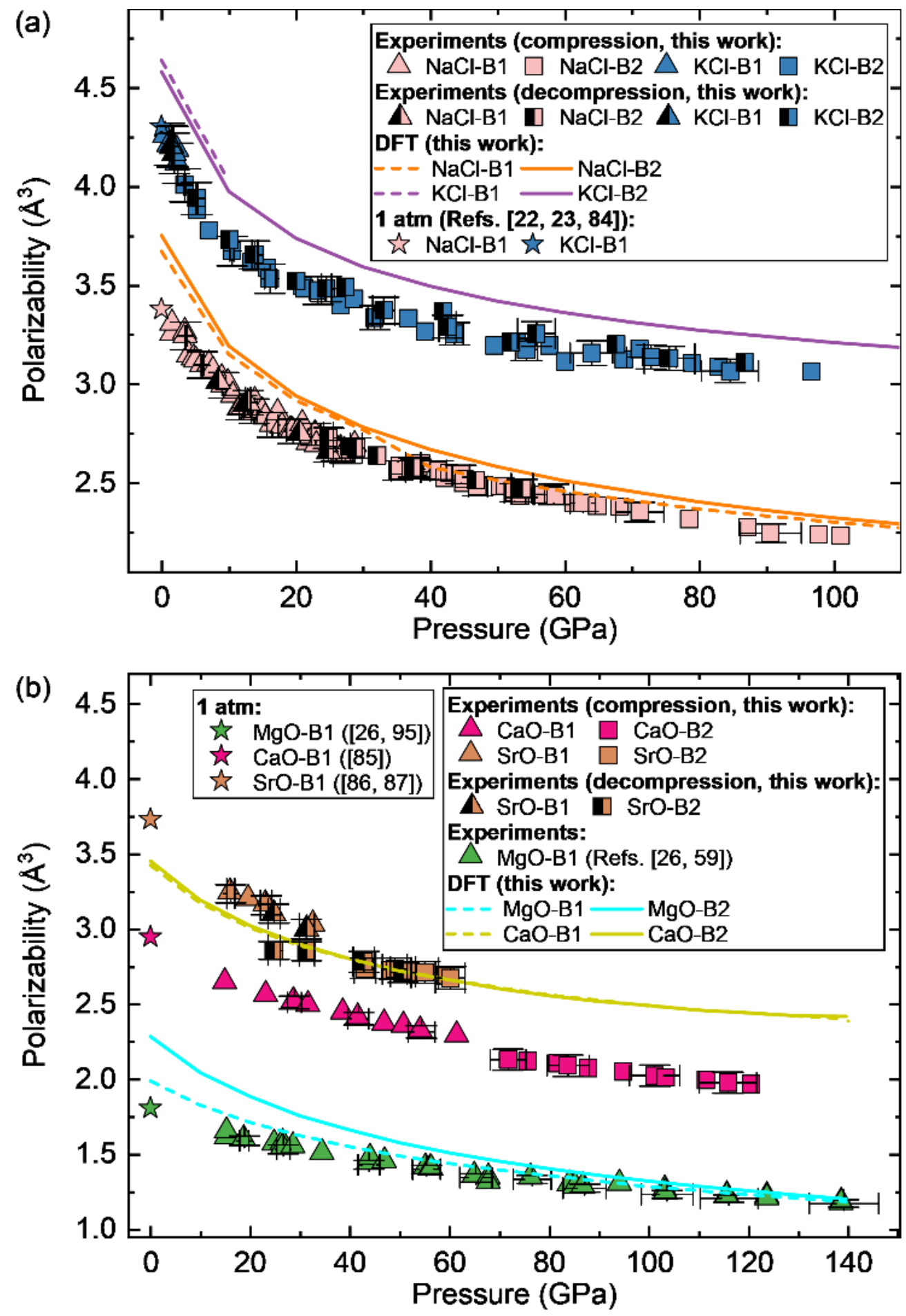

FIG. 3. Electronic polarizabilities of B1 and B2 phases of NaCl, KCl, CaO, SrO, and MgO obtained from the refractive indices in FIG.2 using the Lorentz–Lorenz equation (Eq. 6) and the corresponding equations of state [3-6,14,22,25,87]. The 1 atm values of the refractive indices are from [84-86,95]. Note: for B2 MgO we used our own DFT-PBE calculations of its refractive index and density. Only some error bars are shown because of the high data density. The reproducibility error for the electronic polarizability is ±2-4%, arising from error propagation in the refractive index (~1%) and density (~1-3%). The pressure reproducibility error is ±3% or ±5% depending on the pressure measurement method.

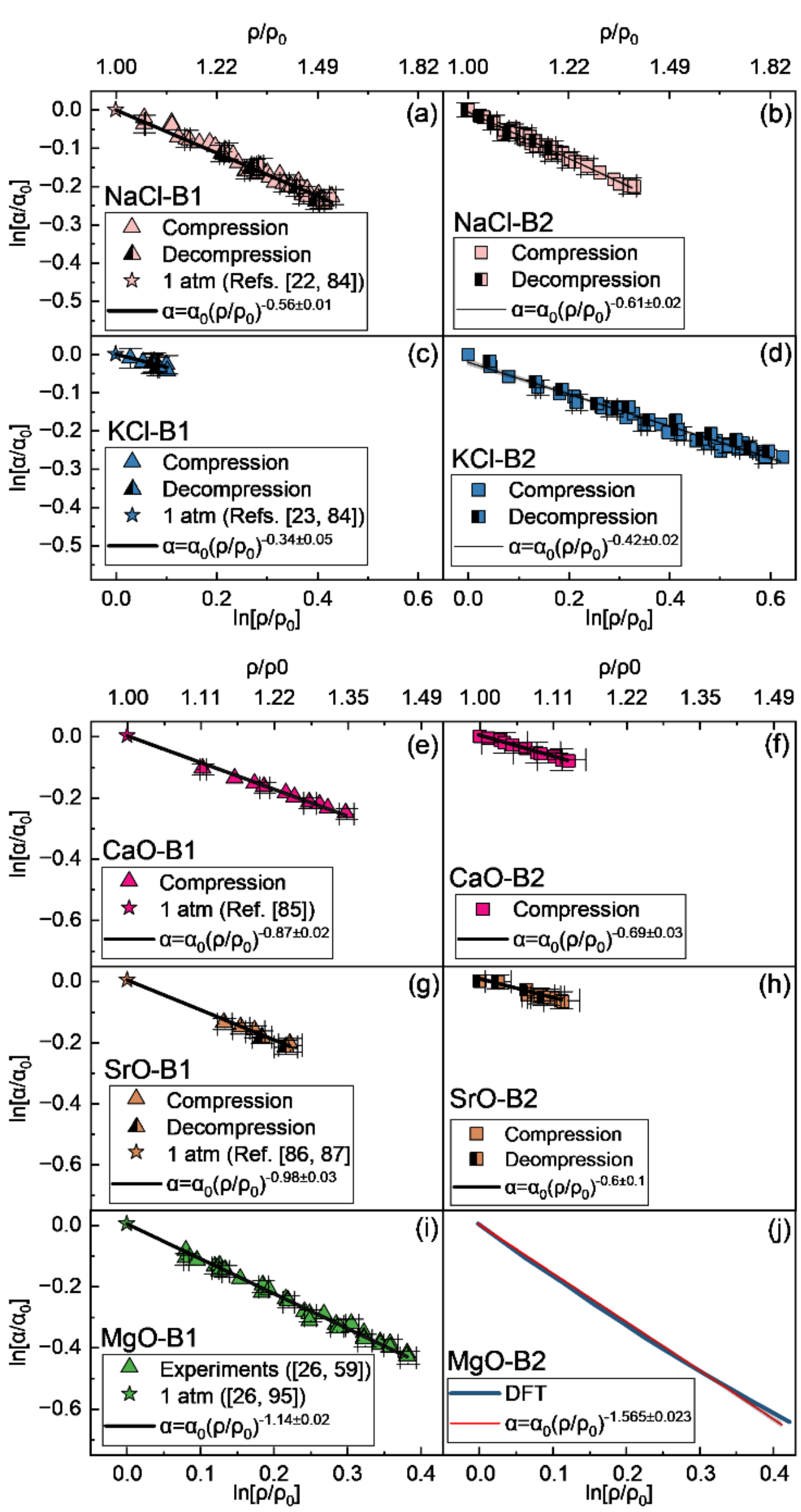

FIG. 4. Relative Lorentz–Lorenz polarizabilities of statically compressed NaCl-B1 (a), NaCl-B2 (b), KCl-B1(c), KCl-B2 (d), CaO-B1 (e), CaO-B2 (f), SrO-B1 (g), SrO-B2 (h), MgO-B1 (i), MgO-B2 based on our DFT-PBE computations (j). The 1 atm values of the refractive indices are from [84-86,95]. The densification calculated using the appropriate EoSs [3,5,6,14,22,25,87]. Lines are linear unweighted fits to the Eq. (8). The uncertainties in the fit parameters indicate the 95% confidence intervals. The uncertainties in the logarithmic values were obtained by error propagation from the relative uncertainties of the non-logarithmic quantities. Only some error bars are shown because of the high data density

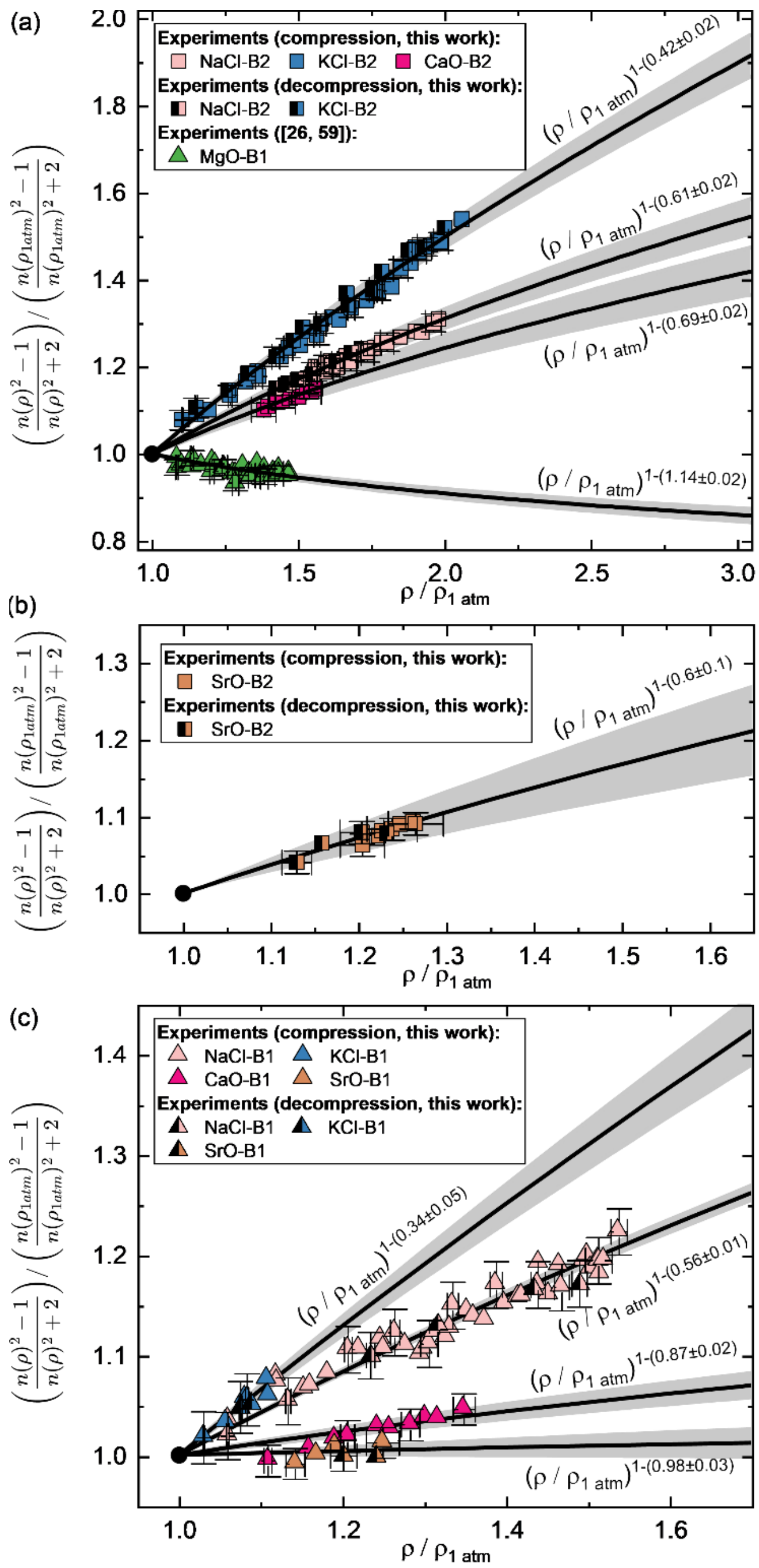

FIG. 5. Normalized Lorentz-Lorenz factors (Eq. 11) for NaCl-B2, KCl-B2, CaO-B2 and MgO-B1 (a), for SrO-B2 (b), and for NaCl-B1, KCl-B1, CaO-B1, and SrO-B1 (c). Lines are Eq. (11) with the strain polarizability parameter from the fits in Fig 4. For phases stable at 1 atm, the normalization was performed using the 1 atm refractive indices [84-86,95]. For the B2 phases, the reference values used for normalization were obtained by extrapolating the fits shown in Fig. 4 to the theoretical densities at 1 atm and then calculating the refractive indices using the Lorentz–Lorenz (LL) equation. The resulting values were n = 1.5982 for NaCl-B2, n = 1.5646 for KCl-B2, n = 1.8640 for CaO-B2, and n = 1.8945 for SrO-B2. The appropriate EoSs are from [3-6,14,22,25,87]. Only some error bars are shown because of the high data density

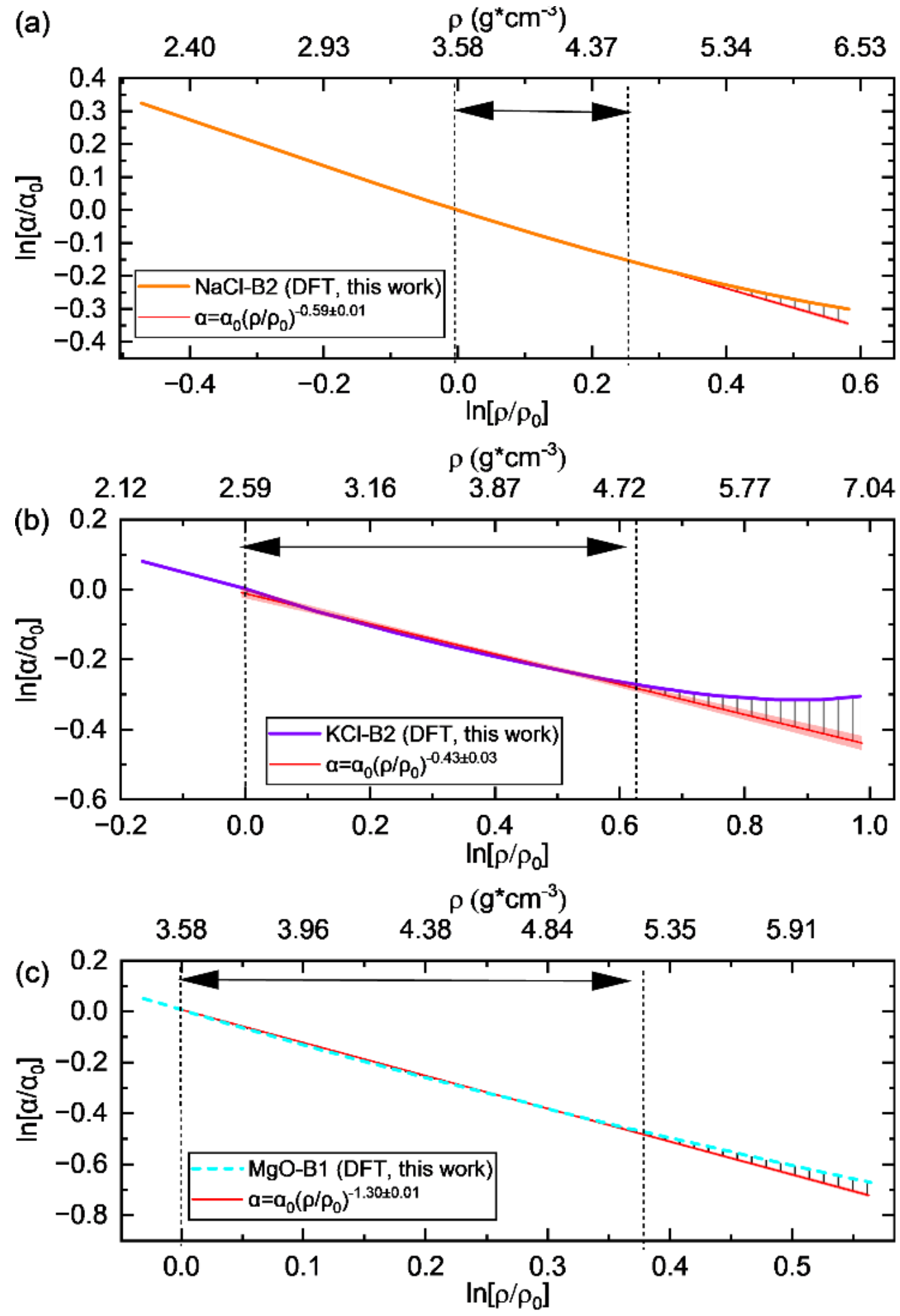

FIG. 6. Lorentz–Lorenz polarizabilities of statically compressed NaCl-B2, with DFT-PBE results shown as an orange line; (b) statically compressed KCl-B2, with DFT-PBE results shown as a purple line; and (c) statically compressed MgO-B1 from [59] with DFT-PBE results shown as a cyan line.. The shaded areas show the deviations of the constant strain polarizability parameter model from the values obtained by DFT-PBE calculations. The DFT-PBE results were normalized to the density of the first experimental data point and to the corresponding DFT-PBE polarizability at that density. The normalized DFT-PBE data were then fitted with a linear function over the same densification range as the experimental data.

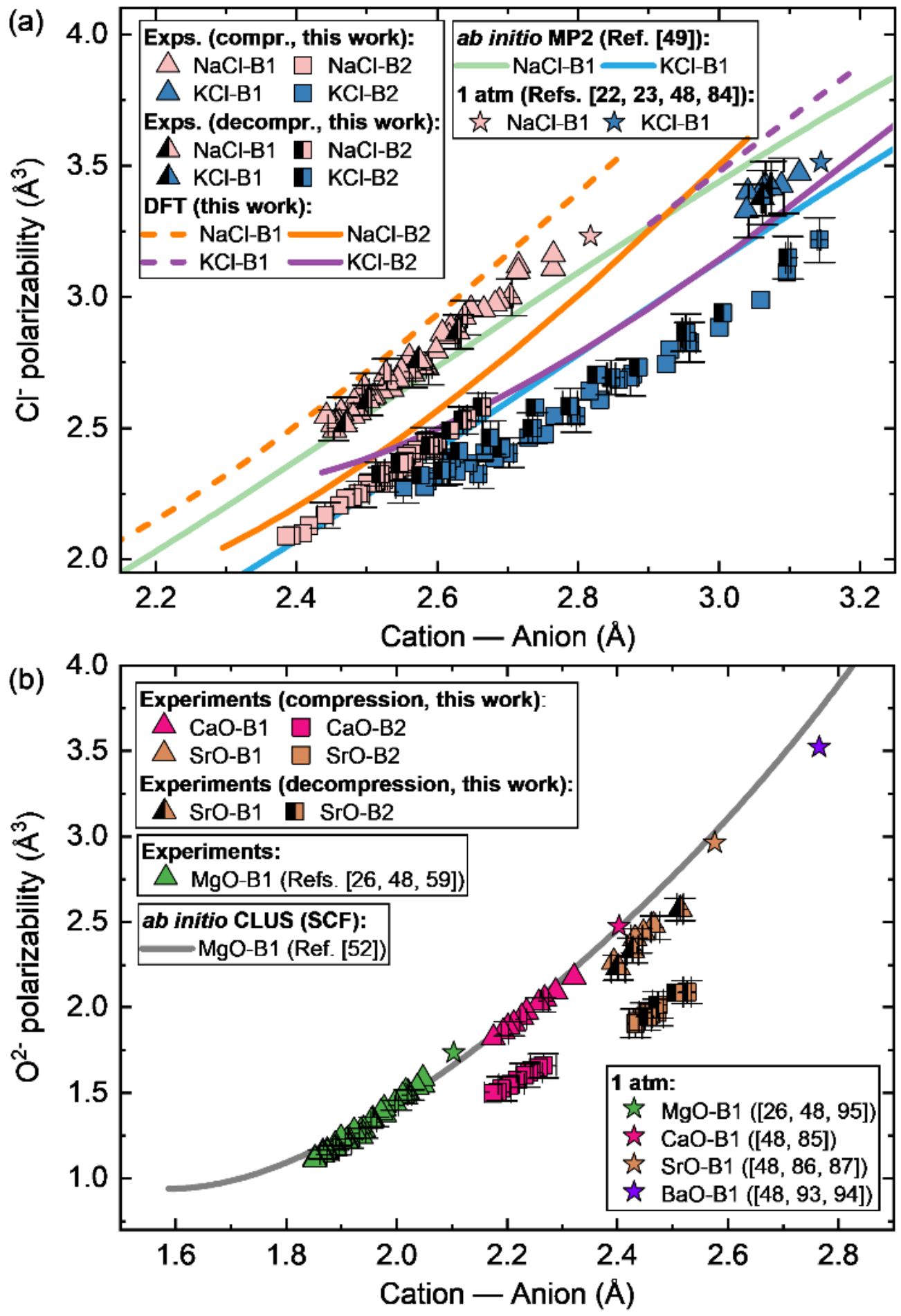

FIG. 7. Polarizabilities of $Cl^-$ (a) and $O^{2-}$ (b) calculated from the Lorentz–Lorenz electronic polarizability by subtracting the recommended cation (in-crystal) polarizabilities [48]. The symbols are our experimental data and data from the previous works, and curves denote either anion polarizabilities obtained from our own DFT-PBE results or from *ab initio* computations reported in the literature [49,52]. The 1 atm values of the refractive indices are from [84-86,93,95]. The cation-anion distance calculated using the appropriate EoSs [3-6,14,22,25,87,94]. Only some error bars are shown because of the high data density. The reproducibility error for the electronic polarizability is ±2-4%, arising from error propagation in the refractive index (~1%) and density (~1-3%). The acronym MP2 stands for second-order Møller–Plesset perturbation theory.